# A revisit to the GNSS-R code range precision

**O. Germain and G. Ruffini**

Starlab, C. de l'Observatori Fabra s/n, 08035 Barcelona, Spain, http://starlab.es
Contact: *olivier.germain@starlab.es*, *giulio.ruffini@starlab.es*

*Abstract* – **We address the feasibility of a GNSS-R code-altimetry space mission and more specifically a dominant term of its error budget: the reflected-signal range precision. This is the RMS error on the reflected-signal delay, as estimated by waveform retracking. So far, the approach proposed by [Lowe *et al.*, 2002] has been the state of the art to theoretically evaluate this precision, although known to rely on strong assumptions (e.g., no speckle noise). In this paper, we perform a critical review of this model and propose an improvement based on the Cramer-Rao Bound (CRB) approach. We derive closed-form expressions for both the direct and reflected signals. The performance predicted by CRB analysis is about four times worse for typical space mission scenarios. The impact of this result is discussed in the context of two classes of GNSS-R applications: mesoscale oceanography and tsunami detection.**

## I. INTRODUCTION

GNSS-R, the use of Global Navigation Satellite Systems (GNSS) reflected signals is a powerful and potentially disruptive technology for remote sensing: wide coverage, passive, precise, long-term, all-weather and multi-purpose. GNSS emit precise signals which will be available for decades as part of an emerging infrastructure resulting from the enormous effort invested in GPS, GLONASS, Galileo and augmentation systems. A key advantage of GNSS-R is its "multistatic" character: unlike monostatic systems, a single receiver will collect information from a simultaneous set of reflection points associated to GNSS emitters. A system in low Earth orbit capable of collecting GPS, Galileo and GLONASS data would potentially be combing the surface with more than a dozen reflection tracks at the same time (for a review, see [Ruffini 2006]).

An important aspect is that GNSS signals are very weak as they were not designed for radar applications; yet they contain a wealth of information. For this reason, signal processing plays an important role. The first detection of GNSS signals from space was documented in [Lowe et al., 2002]. More recently, GPS-R L1 C/A signals have been successfully detected from a dedicated experiment in space using a moderate gain antenna [Gleason *et al.*, 2005], complementing a large number of experiments from aircraft and stratospheric balloons. The resulting data will be used to further validate models.

The reflection process affects the signal in several ways, at the same time degrading (from the point of view of detection) and loading it with information from the reflecting surface. The waveform amplitude is normally reduced, the shape distorted and signal coherence mostly lost.

While GNSS-R cannot provide the precision of dedicated radar altimetry missions, it offers a significant advantage thanks to its multistatic character. The impact of GNSS-R altimetry data to global circulations models has been studied through simulations, with very promising results [Le Traon *et al.*, 2002]. Another recent impact study has focused on the potential of GNSS-R to detect Tsunami's [Martín-Neira *et al.*, 2005]. A dedicated GNSS altimetry system could provide timely warnings, potentially saving many lives. As described in [Soulat *et al.*, 2005], simulations have indicated that a global 100% tsunami detection rate in less than two hours is possible with a ten satellite GNSS-R constellation.

Altimetry in GNSS-R can be carried out in two general ways, depending on the ranging principle used. In code altimetry, our focus here, the code is used for ranging with the direct and reflected signals. In phase altimetry, the phase of the signal is used. All of this is rather similar to normal GNSS processing. The main difference is that the reflected signal is affected by the reflection process, which generally distorts the triangular waveform shape of the return and renders the reflected signal very incoherent. This makes the ranging task rather challenging.

## II. RANGE PRECISION AND ALTIMETRY

Contrarily to classical radar altimetry, range precision is a dominant factor in the error budget for a GNSS-R code-altimetry space mission, due to the much lower modulation bandwidth (1 MHz or 10 MHz for the GPS C/A and P codes respectively). If the direct signal error is considered negligible in front of the reflected signal error, the altimetry precision $\sigma_h$ writes simply as a function of the reflected signal range precision $\sigma_R$:

$$\sigma_h = \frac{\sigma_R}{2\sin\varepsilon}, \qquad \textbf{Eq. 1}$$

where $\varepsilon$ is the transmitter elevation angle. [Lowe *et al.* 2002] proposed a simple approach to assess $\sigma_R$ and since then, the majority of space mission feasibility studies (e.g. the ESA PARIS and STERNA studies) rely on this reference as an approximation. However, this model is known to neglect important aspects—notably speckle—and a re-evaluation of the matter is necessary.

Section III presents a critical review of the state of the art and discusses the model validity. Section IV introduces the

Cramer-Rao Bound (CRB) theory which constitutes the foundation of our analysis approach. This methodology is then applied to both the direct and reflected GNSS signals to derive closed-form expressions of range precision in sections V and VI respectively. Finally, the impact of new performance predictions is illustrated in section VII where mission scenarios are discussed in the light of two classes of applications.

### III. STATE OF THE ART REVIEW

The approach proposed in [Lowe *et al.* 2002] basically assumes that range precision for the reflected signal can be evaluated (to first order) in the same way as for the direct signal. The reflected waveform is assumed to be re-tracked using the algorithm of [Thomas, 1995]. This algorithm estimates the direct waveform's delay using three points (the peak and its two immediate neighbours) to determine the peak sub-sample position. In the limit of low thermal noise the precision of this algorithm turns out to be

$$\sigma_R \approx \frac{1}{\sqrt{2}} \frac{\tau_c}{snr} \sqrt{1 - C(2)} \cdot \qquad \textbf{Eq. 2}$$

where $\tau_c$ is the chip length, $C(2)$ is the correlation factor between amplitudes separated by two lags and *snr* is the signal to noise ratio, defined as the ratio between average and standard deviation of the peak amplitude.

The approach proposed by [Lowe *et al.*] suffers from several limitations. First, it is valid for relatively high SNR only. Second, the derived expression is tied to the choice of a particular estimator. It cannot then be considered applicable to others and as such, it does not address the general case of retracking where an arbitrary number of waveform's points are fit by a model. Third, the derived expression (and associated estimator) assumes a direct signal statistical model whereas the reflected signal is quite different. The waveform's fluctuations are caused by thermal noise but also by speckle. Besides, the waveform's shape is far from the triangular aspect of the direct signal. Finally, the retracking will presumably not be done on the peak of the waveform (which is known to be an unstable and badly localized feature of the reflected signal) but rather on its leading edge.

For these reasons, it appears necessary to re-assess the matter in a more systematic fashion, using appropriate tools from Estimation Theory.

### IV. CRAMER-RAO BOUND

The context of the present problem is Estimation Theory. The CRB methodology allows predicting the best achievable performance in estimation problems for which the stochastic nature of the observation can be described by a probability distribution function (PDF). Formally, the problem comes to estimate a parameter $\theta$ (e.g., the delay) from a random observation $X$ (the complex waveform, a vector), knowing its PDF $p(X,\theta)$. Then, the RMS precision of any non-biased estimator of $\theta$ has a lower bound (see e.g. [Kay 1993]):

$$CRB = \left[ -<\frac{\partial^2}{\partial \theta^2} \log p(X,\theta) > \right]^{-1/2}. \qquad \textbf{Eq. 3}$$

Focusing on complex, vectorial Gaussian-distributed signals, the PDF is given by

$$p(X,\theta) = \frac{1}{\pi^{card(X)}|\Gamma|} \exp\left[-(X-m)^+ \cdot \Gamma^{-1} \cdot (X-m)\right], \qquad \textbf{Eq. 4}$$

where $m=<X>$ and $\Gamma=<(X-m)(X-m)^+>$ are the mean vector and covariance matrix of the complex signal vector $X$ respectively. In this case, the CRB expression is

$$CRB^{-2} = \frac{\partial^2}{\partial \theta^2} \log|\Gamma| + 2\left(\frac{\partial m}{\partial \theta}\right)^+ \Gamma^{-1} \left(\frac{\partial m}{\partial \theta}\right) + \sum_{ij} \Gamma^+_{ij} \frac{\partial^2}{\partial \theta^2} \Gamma^{-1}_{ij} \qquad \textbf{Eq. 5}$$

This expression is the starting point for evaluating the GNSS direct/reflected range precisions, as developed in the two following sections.

### V. DIRECT SIGNAL RANGE PRECISION

The RF signal received by the direct antenna can be seen as an attenuated ($\alpha$ factor) and delayed (by $\theta$) version of the GNSS code $C$ emitted by the transmitter, and corrupted by additive thermal noise $\sigma b$ (where $b$ is a complex zero-mean unit-variance white-noise Gaussian random process and $\sigma$ a real scaling factor). The waveform is produced by correlating this input signal with a clean replica of the GNSS down-converted signal, leading to the complex waveform

$$X = [\alpha \cdot C_\theta + \sigma \cdot b] \otimes C, \qquad \textbf{Eq. 6}$$

defined along the time-delay axis $\tau_i$ (i.e. the correlation lag vector). Introducing the GNSS code autocorrelation function,

$$\chi \equiv C \otimes C, \qquad \textbf{Eq. 7}$$

it is immediate to write expressions for the mean complex waveform and its covariance matrix:

$$m_i = \alpha \cdot \chi\left(\frac{\tau_i - \theta}{\tau_c}\right), \qquad \textbf{Eq. 8}$$

$$\Gamma_{ij} = 2\sigma^2 \chi\left(\frac{\tau_i - \tau_j}{\tau_c}\right). \qquad \textbf{Eq. 9}$$

Having a Gaussian-distributed signal allows to use Eq. 5 and plugging the mean and covariance expressions leads to the CRB for the direct signal delay estimation, that is, the best possible performance for direct signal range precision:

$$CRB^{-2} = (2 - \frac{\pi}{2}) \frac{SNR_1^2}{\tau_c^2} \sum_{ij} \partial \chi\left(\frac{\tau_i}{\tau_c}\right) \partial \chi\left(\frac{\tau_j}{\tau_c}\right) \chi^{-1}_{ij}, \qquad \textbf{Eq. 10}$$

where we have introduced the one-shot thermal $SNR_1$, defined as the ratio between the mean amplitude of the peak to the thermal noise amplitude STD (the so-called, "grass fluctuations"):

$$SNR_1 \equiv \frac{\alpha}{\sigma\sqrt{2-\pi/2}}. \qquad \textbf{Eq. 11}$$

Note that for the direct signal this SNR definition can be linked to the previous one,

$$SNR_1 \approx \sqrt{2} snr.$$  **Eq. 12**

The CRB expression can now be compared to the state of the art model. For this purpose, Eq. 10 should be further simplified by adopting the assumptions that $\chi$ is a triangle function and that only three points of the waveform are retained for retracking (the peak and its two immediate neighbours). Doing this, we recover Eq. 2. This exercise illustrates the strength of the CRB approach for deriving generic performance expression adaptable to a particular algorithm and also proves that the Thomas estimator is an efficient one (i.e. reaching its Cramer-Rao bound) in the limit of high enough SNR.

Figure 1 gives values of the direct signal (GPS C/A code) range precision as a function of 1/SNR (i.e. NSR). The waveform is sampled from -300m to 300m with a step of 15m (i.e. 20 MHz). As expected, the CRB approach is in full agreement with the state of the art model. To further validate these results, we have performed Monte-Carlo simulations. Realizations of the model of Eq. 6 have been re-tracked using two estimators: the Maximum Likelihood Estimator (MLE), known to be efficient, and the Thomas algorithm. MLE results match well to the theoretical CRB, except for very low SNR where a slight departure is observed. As expected, the Thomas algorithm is efficient for high SNR but deviates from optimality at severe noise levels.

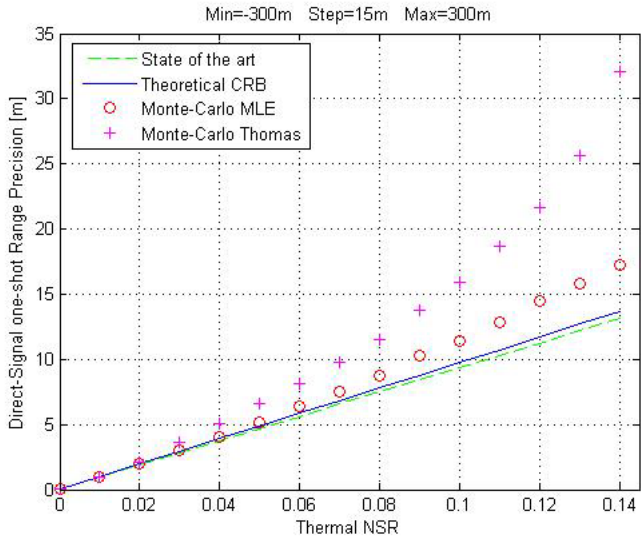

**Figure 1 - Direct signal range precision vs. one-shot thermal NSR, given by the state of the art model (Eq. 2), the CRB approach (Eq. 10) and Monte-Carlo simulations conducted with the MLE and Thomas estimators.**

## VI. REFLECTED SIGNAL RANGE PRECISION

The expression for the complex reflected waveform involves two contributions: one is the GNSS electric field scattered by sea-surface and the other is thermal noise, as for the direct signal,

$$X = (\alpha \cdot U + \sigma \cdot b) \otimes C = \alpha \cdot u + \sigma \cdot b \otimes C,$$  **Eq. 13**

where $U$ is the scattered electric field and $u$ the electric field after correlation with a signal replica. From space and for the majority of sea-states, it can reasonably be assumed that the sea-surface scattering contribution follows fully-developed speckle statistics, that is, a complex, vectorial, zero-mean, Gaussian PDF. Since thermal noise is also Gaussian, the reflected complex waveform is Gaussian distributed with parameters

$$m_i = 0,$$  **Eq. 14**

$$\Gamma_{ij} = <u_i u_j^*> + \frac{\pi}{4-\pi} \frac{1}{SNR_1^2} \chi\left(\frac{\tau_i - \tau_j}{\tau_c}\right).$$  **Eq. 15**

The CRB expression immediately follows:

$$CRB^{-2} = \frac{\partial^2}{\partial \theta^2} \log|\Gamma| + \sum_{ij} \Gamma_{ij}^+ \frac{\partial^2}{\partial \theta^2} \Gamma_{ij}^{-1}.$$  **Eq. 16**

The tricky part is now to evaluate the covariance of the scattered filtered field $<u_i.u_j^*>$. The starting point is the EM integral equation of [Zavorotny and Voronovich, 2000] modelling the scattered filtered field $u$. Now, we emphasize that the critical feature for our purpose is the waveform leading edge which is obtained by integration of sea-surface scatterers in the vicinity of the specular point. In this regime and from space, the signal covariance is largely dominated by the radar ambiguity function, i.e. by the GNSS autocorrelation. In other words, we assume that the antenna pattern and the glistening zones are much larger than the first-chip zone. In addition, we simplify further the study by limiting ourselves to reflections occurring at nadir. Under these assumptions, the covariance of the scattered filtered field, in the leading edge regime, simplifies to

$$<u_i u_j^*> = \int_0^\infty \chi\left[\frac{\tau_i - \theta - \xi}{\tau_c}\right] \cdot \chi\left[\frac{\tau_j - \theta - \xi}{\tau_c}\right] d\xi.$$  **Eq. 17**

Figure 2 illustrates the $\Gamma$ covariance matrix. We highlight again that this model is acceptable for the description of the leading edge but cannot render the behaviour of the waveform's trailing edge (which is affected by the finite size of antenna beam and glistening zone).

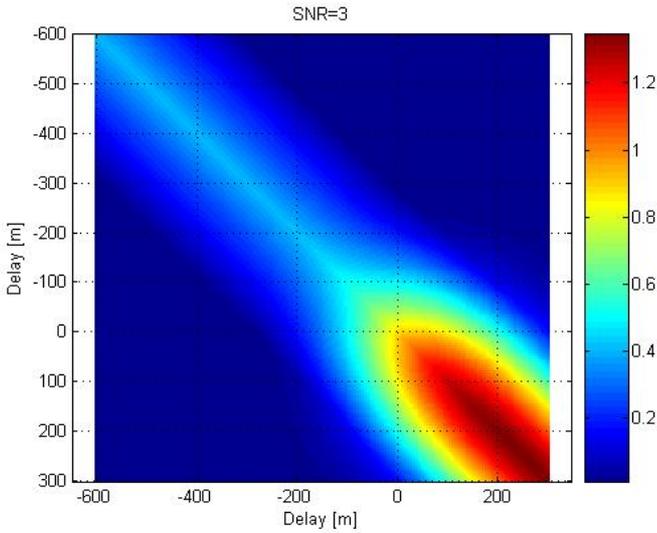

**Figure 2 - Illustration of the reflected-signal covariance matrix model (Eq. 15), obtained for SNR=3.**

Figure 3 provides values for the reflected signal (GPS C/A code) range precision as a function of NSR. The waveform is sampled from -400m to 300m with a step of 15m (i.e., ~20 MHz). The re-assessment leads to more pessimistic results than in previous analyses: typically, the range precision computed with the CRB approach is predicted ~4 times worse. Besides, it is worth noting that the asymptotic range precision for infinite SNR is now predicted finite. Even without thermal noise (e.g., with a very large antenna), the waveform is still degraded by speckle and this remains a limitation for delay estimation.

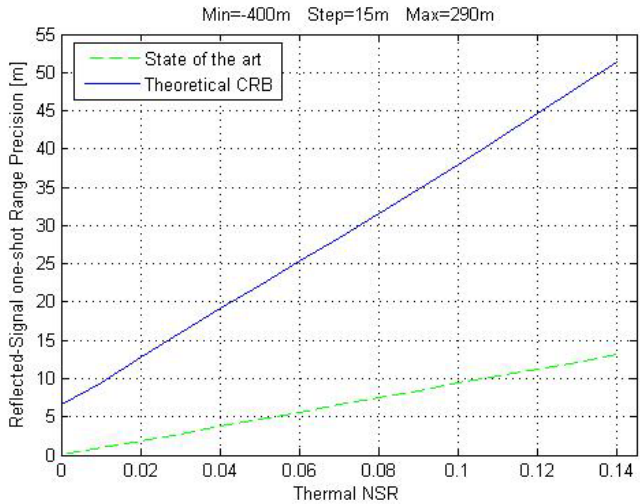

**Figure 3 - Reflected signal range precision vs. one-shot thermal NSR, given by the state of the art model (Eq. 2), the CRB approach (Eq. 16).**

## VII. ALTIMETRY SCENARIO STUDY

The impact of this result is now discussed. A simple error budget is assessed for two generic space missions and compared to the requirements of space altimetry applications potentially suitable for GNSS-R. The two proposed missions receive the GPS C/A code and are characterized by their altitude (500 or 700 km) and antenna gain (28 or 34 dB). A link budget model developed elsewhere [CNES ALT GNSSR, 2006] allows computing the expected thermal SNR and the coherence time of the reflected signal, which is needed to compute the number of independent samples in one second. The altimetric precision is then derived according to Eq. 1.

| Parameter | Mission 1 | Mission 2 |
|---|---|---|
| Altitude (km) | 500 | 700 |
| Antenna Gain (dB) | 28 | 34 |
| Waveform sampling step (m) | 15 | 15 |
| One-shot thermal SNR (linear) | 12 | 22 |
| Coherence time (ms) | 0.8 | 0.9 |
| One-shot nadir range precision (m) | 32.6 | 20.8 |
| One.sec nadir range precision (cm) | 92 | 62 |
| One-sec nadir altimetric precision (cm) | 46 | 31 |

**Table 1 - Performance of two GNSS-R space missions using the GPS C/A code (nadir case).**

Table 2 shows user requirements for mesoscale oceanography and tsunami detection, expressed as the altimetric and spatial scales of signatures to be observed. The two missions using the C/A code meet the requirements of strong tsunami detection but not the ones of mesoscale oceanography.

| Parameter | Mesoscale oceanography | Strong tsunami detection |
|---|---|---|
| Altimetric scale (cm) | 5 | 20 |
| Spatial scale (km) | 100 | 100 |
| Allowed integration time (s) | 13.3 | 13.3 |
| 1sec altimetry precision (cm) | 18 | 73 |

**Table 2- User requirements for applications addressed by GNSS-R altimetry.**

The same exercise has been conducted for a GNSS code with a ten times broader bandwidth, namely the GPS P code (Table 3). The performance improvement is rather clear and becomes now compatible with the requirements of mesoscale oceanography.

| Parameter | Mission 1 | Mission 2 |
|---|---|---|
| Altitude (km) | 500 | 700 |
| Antenna Gain (dB) | 28 | 34 |
| Waveform sampling step (m) | 1.5 | 1.5 |
| One-shot thermal SNR (linear) | 4.8 | 8.7 |
| Coherence time (ms) | 2.5 | 2.8 |
| One-shot nadir range precision (m) | 7.7 | 4.3 |
| One.sec nadir range precision (cm) | 38 | 23 |
| One-sec nadir altimetric precision (cm) | 19 | 11 |

**Table 3 - Performance of two GNSS-R space missions receiving the GPS P code (nadir case).**

## VIII. CONCLUSIONS

In this paper, we have carried out a critical review of the state of the art model for GNSS-R range precision. The goal was to revisit the baseline assumption (known to be incorrect) that reflected and direct signals can be treated the same. A rigorous evaluation of the problem, based on the Cramer-Rao Bound methodology, has been conducted. For the direct signal, we have obtained results in agreement with the state of the art, as expected. For the reflected signal, we have shown that precision degrades, as suspected. For instance, a mission receiving the C/A code at 500km with 28dB gain would have a 1-second range precision of 1m at nadir. This is due to the impact of speckle noise and the shape change in the reflected signal.

These results question the suitability of a C/A code GNSS-R mission focusing on mesoscale altimetry. The use of 1-MHz codes (e.g. GPS C/A) remains acceptable to detect strong tsunamis (20 cm over 100 km) but mesoscale oceanography (5 cm over 100 km) would be realistic only with 10-MHz codes (e.g., GPS P code).

The availability of such signals as well as those with an even higher bandwidth (up to 50 MHz with the E5 signal) provided by the European Galileo system will further increase the potential of this technique [Galileo OS SIS ICD, 2006].

Future work should consolidate these results with more numerical simulations and experimental validation either using space data or adapting the model to low altitudes and take benefit of available airborne/coastal data. Finally, an in-depth study of the Galileo signal structure impact on GNSS-R is a very important future research line.

## ACKNOWLEDGMENTS

Part of this work was carried out under CNES contract. We thank C. Tison (CNES) for publication permission.

## REFERENCES

**[CNES ALT GNSSR, 2006]** « Evaluation des Performances Altimétriques à partir de Signaux GNSS », CNES contract, 2005/2006.

**[Galileo OS SIS ICD, 2006]** Galileo Open Service, Signal in Space Interface Control Document, Draft 0, May 2006.

**[Gleason et al., 2005]** "Detection and Processing of Bi-Statically Reflected GPS Signals From Low Earth Orbit for the Purpose of Ocean Remote Sensing", IEEE Trans. on Geoscience and Remote Sensing.

**[Kay, 1993]** "Fundamentals of Statistical Signal Processing – Estimation Theory", Prentice Hall, Upper Saddle River, NJ.

**[Le Traon et al., 2002]** "Mesoscale Ocean Altimetry Requirements and Impact of GPS-R measurements for Ocean Mesoscale Circulation Mapping" Abridged Starlab ESA/ESTEC Technical Report, available at http://arxiv.org/abs/physics/0212068.

**[Lowe et al., 2002]** "First spaceborne observation of an Earth-reflected GPS signal", Radio Science 37(1).

**[Martin-Neira et al., 2005]**, "Detecting tsunamis using PARIS concept", URSI Conf. on Microwave Remote Sensing of the Earth, Ispra, Italy, 20-21 April 2005.

**[Ruffini, 2006]** IEEE GRS March Newsletter, p. 15-21.

**[Soulat et al., 2005]** "PARIS mission impact analysis", GNSSR'05 workshop, Surrey, UK, June 2005.

**[Thomas, 1995]** "Signal Processing Theory for the Turbo Rogue Receiver", JPL Publication 95-6.

**[Zavorotny and Voronovich, 2000]** "Scattering of GPS signals from the ocean with wind remote sensing application", IEEE Trans. Geoscience and Remote Sensing, 38(2):951-964.